\documentclass[aps,twocolumn,prd,showpacs,nofootinbib]{revtex4}
\usepackage{amsmath}
\usepackage{graphicx}
\usepackage{dcolumn}
\usepackage{bm}
\usepackage{amssymb}
\usepackage{latexsym}

\def\be{\begin{equation}}
\def\ee{\end{equation}}
\def\ba{\begin{eqnarray}}
\def\ea{\end{eqnarray}}

\begin{document}

\title{
Entropy of Microwave Background Radiation in Observable Universe }

\author{Yun-Song Piao}
\affiliation{
College of Physical Sciences, Graduate School of
Chinese Academy of Sciences, YuQuan Road 19A, Beijing 100049,
China}

\begin{abstract}

We show that the cosmological constant at late time places a bound
on the entropy of microwave background radiation deposited in the
future event horizon of a given observer, $S\leq
S_{\Lambda_0}^{3/4}$. This bound is independent of the energy
scale of reheating and the FRW evolution after reheating. We also
discuss why the entropy of microwave background in our observable
universe has its present value.

\end{abstract}
\pacs{98.80.Cq}

\maketitle

Recent observations have indicated that the universe is
accelerating, which suggests that the universe might be being
dominated by a dark energy component with state equation $\kappa
=p/\rho <-1/3$, see Refs. \cite{PR, C2, S, ZF, CST} for reviews on
various aspects. The simplest form of dark energy is a small and
positive cosmological constant, which can fit data very nicely and
is phenomenologically one of the most appealing choices so far.
The universe with a cosmological constant will generally
asymptotically approach a dS equilibrium and any observer in it
will eventually be surrounded by an event horizon. This limits the
region of universe accessible to any given observer, see Refs.
\cite{HKS, Banks, DKS, DV, Piao} for some discussions on
interesting issues.

Recently, Fischler et.al. \cite{FLP} have pointed out that in a
universe dominated by a positive cosmological constant at late
time there is a bound on the total entropy of radiation in the
entire universe, whose scales like $S_{max}\sim \Lambda_0^{-3/4}$,
where $\Lambda_0$ is the value of cosmological constant observed.
This is parametrically smaller than the entropy of dS,
$S_{\Lambda_0} \sim \Lambda_0^{-1}$. Further Banks and Fischler
\cite{BF} argued that this entropy bound places a upper limit on
the total efoldings number ${\cal N}_{tot}$ of inflation that can
be described by a conventional quantum field theory, see also
related Refs. \cite{C}. However, in subsequent discussion in Ref.
\cite{LM} based on the covariant entropy bounds \cite{B}, see also
Refs. \cite{EL}, it has been shown that no bound on the total
${\cal N}_{tot}$ was found. The reason is that the existence of a
horizon only constrains the entropy visible to a given observer on
any homogeneous spacelike slice to be less than the area of its
interaction with the past light cone of the observer, but can not
restrict the entropy outside of the causal future of this given
observer. However, there is actually a bound ${\cal N}_{obs}$ on
the efoldings number that will ever be observable to a given
observer at late time \cite{KKS}. If the observable universe
accelerates forever, the observer will never see more efoldings
number than ${\cal N}_{obs}$. Thus similarly, it may be expected
that there should also be a bound on the entropy of microwave
background radiation deposited in the future event horizon of a
given observer. In this brief report, we will show that this bound
is $S\leq S_{\Lambda_0}^{3/4}$, independent of the energy scale of
reheating and the FRW evolution after reheating.

We assume that the inflation ended at some time $t_e$. The
efoldings number ${\cal N}$ corresponding to the present Hubble
scale is given by \be {\cal N}\equiv \ln{\left({a_eh_e\over
a_kh_k}\right)} \equiv \ln{\left({a_eh_e\over a_0h_0}\right)}
\label{n}\ee  where $h\equiv {\dot a}/a$, and the subscript `$k$'
labels the time of inflation corresponding the present Hubble
scale and `0' is the present time, In an universe without the
future event horizon, a patient observer would be able to see
arbitrarily far back to inflation, since $a_0h_0$ can be made
arbitrarily small at late time. However, the existence of event
horizon $1/h_0$ constrains the size $r$ of the reheating surface
visible to any given observer at late time. Taking (effectively)
$a_0\simeq 1/h_0$, we can obtain this size $r=a_e \simeq
h_e^{-1}e^{\cal N}$. Thus in Planck unit, the entropy $S$ released
into this region of the reheating surface is simply the product of
the entropy density and the volume, \be S\simeq\sigma r^3 \simeq
\rho_e^{1/(1+\kappa)} h_e^{-3} e^{3\cal N} \label{si}\ee where we
have assume that the observable universe is filled with fluid
$p=\kappa \rho$ after reheating and $\sigma \sim
\rho^{1/(1+\kappa)}$ has been used. We also assume that the
reheating after inflation is perfectly efficient, thus the
reheating energy scale $\rho_e$ can approximately equal to the
Hubble scale $\sqrt{h_e}$ during inflation, and also hereafter
there are not other reheating processes bringing a large number of
entropy, such as the decay of some extra productions. During
following FRW evolution, if the observable universe is still
dominated by this kind of fluid, we have $\rho \sim
1/a^{3(1+\kappa)}$. Thus from (\ref{n}), we obtain
the efoldings number \be e^{\cal N} ={a_eh_e \over
a_0h_0}=\left({\Lambda_0\over \rho_e}\right)^{1\over 3(1+\kappa)}
\left({\rho_e\over \Lambda_0}\right)^{1/ 2} =\left({\rho_e\over
\Lambda_0}\right)^{1+3\kappa\over 6(1+\kappa)}\label{en}\ee where
$\rho_0\simeq \Lambda_0$ has been taken. Thus instituting it into
(\ref{si}), the entropy in observable universe can be given by \be
S \simeq \Lambda_0^{-{1+3\kappa\over 2(1+\kappa)}} \label{sii}\ee
Thus it seems that the capacity of store information in the event
horizon at late time only depends on the state equation of fluid
filling it. For the state equation $\kappa =1$, such as black hole
gas \cite{BFI}, $S\simeq \Lambda_0^{-1}\simeq S_{\Lambda_0}$. For
the radiation $\kappa =1/3$, the entropy is \be S \simeq
\Lambda_0^{-3/4} \simeq S_{\Lambda_0}^{3/4}\label{sbound}\ee In
Ref. \cite{BF}, it was pointed out that to avoid a big crunch the
total entropy of microwave background radiation in a system should
be bounded from above by (\ref{sbound}). However, here we point
out that what (\ref{sbound}) bound is only the entropy deposited
in the region of event horizon at late time, which is accessible
to any given observer in future event horizon. Taking the present
observed value of cosmological constant, $\Lambda_0 \sim
10^{-123}$, we have $S\simeq 10^{91}$. This is the maximal entropy
which the microwave background radiation deposited in the event
horizon potentially approaches.

We discuss some interesting cases for further arguments in the
following. The observable universe actually consists of radiation
and matter. The energy density of matter will exceed that of
radiation at some time $t_{eq}$. From (\ref{n}) and (\ref{en}), we
have \be e^{\cal N} ={a_eh_e\over a_{eq}h_{eq}}\cdot
{a_{eq}h_{eq}\over a_0h_0}=\left({\rho_e\over
\rho_{eq}}\right)^{1/ 4}\left({\rho_{eq}\over
\Lambda_0}\right)^{1/ 6} \label{enn}\ee where the subscript `$eq$'
labels the time of matter-radiation equality. Thus from
(\ref{si}), we can obtain \ba S & \simeq & \rho_e^{3/ 4}\cdot
\left({\rho_e\over \rho_{eq}}\right)^{3/ 4}\left({\rho_{eq}\over
\Lambda_0}\right)^{1/ 2}/\rho_e^{{3/ 2}}\nonumber\\
&\simeq & S_{\Lambda_0}^{{3/ 4}}\left({\Lambda_0\over
\rho_{eq}}\right)^{1/ 4}\leq S_{\Lambda_0}^{3/ 4}\label{siii}\ea
In our observable universe, $\Lambda_0/\rho_e\sim 10^{-12}$, thus
we have $S\sim 10^{88}$, which is just the present entropy of
microwave background radiation. This gives a simple explanation
why the entropy of microwave background in our observable universe
is several orders of magnitude lower but not far lower than the
entropy bound (\ref{sbound}). The reason is that the
matter-dominated universe only began in the not far past. The
higher the energy density of matter is, the earlier it will
dominate the universe and thus the larger $\rho_{eq}$ is. Thus
the entropy of microwave background in observable universe will be
lower. The limit case is $\rho_{eq}\simeq \rho_e$, i.e. the
universe just entered into the matter-dominated phase shortly
after reheating, in which the observable universe will has lowest
radiation entropy deposited in the future event horizon of a given
observer. Whereas the higher the energy density of radiation is,
the smaller $\rho_{eq}$ will be, thus the higher the radiation
entropy will be. The limit case is $\rho_{eq}\simeq \Lambda_0$,
i.e. the radiation still dominated the universe up to date, which
corresponds to saturate the entropy bound (\ref{sbound}), i.e.
$S\sim 10^{91}$.

We may suppose that the observable universe can be filled with
some other fluid after reheating and will be still dominated by it
before entering into the radiation-dominated phase. This
corresponds to \ba e^{\cal N} & = & {a_eh_e\over
a_{eq^\prime}h_{eq^\prime}}\cdot {a_{eq^\prime}h_{eq^\prime}\over
a_{eq}h_{eq}}\cdot {a_{eq}h_{eq}\over a_0h_0}\nonumber\\
& = & \left({\rho_e\over \rho_{eq^\prime}}\right)^{1+3\kappa\over
6(1+\kappa)} \left({\rho_{eq^\prime}\over \rho_{eq}}\right)^{1/ 4}
\left({\rho_{eq}\over \Lambda_0}\right)^{1/ 6} \label{en1}\ea
where the subscript `$eq^\prime$' labels the time of
fluid-radiation equality. Note that at time $t_{eq^\prime}$ the
radiation begins to dominate the universe and have the energy
density $\rho_{eq^\prime}$. Thus we can back to the
reheating surface and obtain an equivalent energy density
$\rho_{r}$ of radiation at the time $t_e$ \be {\rho_{r}\over
\rho_{eq^\prime}}\simeq \left({\rho_e\over
\rho_{eq^\prime}}\right)^{4\over 3(1+\kappa)}\label{s1}\ee Thus
instituting (\ref{en1}) and (\ref{s1}) into (\ref{si}),
we can obtain \ba S & \simeq & \rho_{r}^{3/ 4}\cdot
\left({\rho_e\over \rho_{eq^\prime}}\right)^{1+3\kappa\over
2(1+\kappa)}\left({\rho_{eq^\prime}\over \rho_{eq}}\right)^{3/
4}\left({\rho_{eq}\over
\Lambda_0}\right)^{1/2}/ \rho_e^{{3/ 2}}\nonumber\\
&\simeq & S_{\Lambda_0}^{3/ 4}\left({\Lambda_0\over
\rho_{eq}}\right)^{1/ 4}\label{siv}\ea This
is the same as (\ref{siii}). Thus the intervening of other fluid
phase before the radiation-dominated phase does not affect our
result.

We can also forward the entropy of microwave background radiation
to the future and note that after the time $t_0$, the observable
universe will enter into a dS phase, thus we have \be e^{{\cal
N}^{\prime}} ={a_eh_e\over a_{eq}h_{eq}}\cdot {a_{eq}h_{eq}\over
a_0h_0}\cdot {a_0h_0\over ah_0}=e^{{\cal N}-{\cal
N}(t)}\label{nv}\ee where (\ref{enn}) has been used and ${\cal
N}(t)=h_0(t-t_0)$. We can see that the comoving Hubble scale $ah$
begins to grow, as in early inflation, which results in the
decrease of the effective efoldings number ${\cal N}^{\prime}$,
which is the efoldings number visible to any given observable at
late time of $t_0$ \cite{KKS}, and thus the entropy $S$ of
microwave background radiation. Let us see this case further. Note
that at the time $t_f-t_0={\cal N}/h_0$, the comoving Hubble scale
equals its value at reheating, thus we have ${\cal N}^{\prime}
\simeq 0$. This means that all perturbations including very last
perturbation generated during inflation will be pushed back out of
the horizon again. Instituting ${\cal N}^\prime$ at the time $t_f$
into (\ref{si}), we obtain $S\simeq \rho_e^{-3/4}$, in which
$\kappa = 1/3$ has been taken. This is just $S_{inf}^{3/4}$ of dS
entropy during inflation. Note that \be S\sim h_0^{-3}
T^3\label{sht}\ee we can obtain the temperature $T_f\sim
h_0\rho_e^{-1/4}$ of microwave background at the time $t_f$. This
temperature is still far larger than the characteristic
temperature $T_h=h_0/2\pi$ of event horizon \cite{GW}. However,
since the observable universe will accelerate forever, the
temperature of microwave background can be eventually redshift to
a point where $T \sim h_0$ and the noise of Hawking radiation will
begin to overwhelm the microwave background. From (\ref{sht}), we
have the eventual radiation entropy $S\sim 1$. This means that
after this time it would be impossible to extract any information
about early universe from the microwave background.

The above discussions also apply to the case that the current and
future evolution is dominated by phantom, in which $\kappa <-1$,
and thus a well-defined event horizon exists. During
phantom-dominated the energy density of phantom is increased with
the time. This leads that the scale of event horizon decreases
continuously, and from (\ref{sbound}), thus the value of entropy
bound on the microwave background radiation deposited in the event
horizon. This result is different from the constant entropy bound
with the cosmological constant. The entropy bound varies
monotonously makes us able to expect that there should be a
maximal entropy bound, which can be seen as follows. The
phantom-dominated evolution only begins at present, and its energy
density is determined by current observations and is approximately
$\Lambda_0\sim 10^{-123}$. Thus from (\ref{sbound}), we have the
entropy bound $S\sim 10^{91}$ on the microwave background
radiation deposited in ``present" event horizon. Now we forward
the entropy bound to the future and have \be e^{{\cal N}^{\prime}}
={a_eh_e\over a_{eq}h_{eq}}\cdot {a_{eq}h_{eq}\over a_0h_0}\cdot
{a_0h_0\over ah}=e^{{\cal N}-{\cal N}(t)}\label{nvv}\ee where
(\ref{enn}) has been used. Reconsidering Eq. (\ref{en}), we have
\be e^{{\cal N}(t)}={ah\over a_0h_0}=\left({\rho\over
\Lambda_0}\right)^{1+3\kappa\over 6(1+\kappa)} \ee where $\rho$ is
the energy density of phantom. We can see that since $\kappa <-1$
and $\rho
> \Lambda_0$, ${\cal N}(t)$ is always positive. Thus instituting
${\cal N}^\prime$ of (\ref{nvv}) to (\ref{si}), we can find that
the bound of entropy on the microwave background radiation
deposited in the event horizon is decreased in the future, and
thus the bound at present is the maximal entropy bound.

In summary, we have shown that the cosmological constant at late
time places a bound on the entropy of microwave background
radiation deposited in the future event horizon of a given
observer, $S\leq S_{\Lambda_0}^{3/4}\sim 10^{91}$, which is
independent of the energy scale of reheating and the FRW evolution
after reheating. However, this dose not means that there is a
limit on the total radiation entropy generated after inflation,
since if inflation lasts long enough, the total entropy may exceed
greatly the above bound. In fact due to the presence of
cosmological constant, not all regions of reheating surface lie
inside the causal patch of a given late-time observer, thus the
entropy deposited in the future event horizon of this observer is
only a portion of total entropy, which it is that obeys our bound.
The entropy of microwave background in our observable universe is
$S_{\Lambda_0}^{3/ 4}(\Lambda_0/ \rho_{eq})^{1/ 4}\sim 10^{88}$
since $\Lambda_0\sim 10^{-12} \rho_{eq}$. The reason that it seems
not too far lower than the bound value is that the
matter-dominated phase only began in the not too far past. This
work might bring a litter insight why our observable universe look
like so.

\textbf{Acknowledgments} This work is supported in part by NNSFC
under Grant No: 10405029, 90403032 and also in part by National
Basic Research Program of China under Grant No: 2003CB716300.

\end{document}